# Hypolipidemic effect of brown seaweed (*Sargassum crassifolium*) extract in vivo (Study of histopathology, mRNA expression, and immunohistochemistry (IHC) with VCAM-1, ICAM-1, and MCP-1 parameters)

**Sarah Zaidan[1]\*, Syamsudin Abdillah[1], Nur Arfian[2], Wawaimuli Arozal[3]**

Corresponding email: sarah.zaidan@univpancasila.ac.id

[1]Faculty of Pharmacy, Universitas Pancasila, Srengseng Sawah, Jagakarsa, South Jakarta, Indonesia 12640.

[2]Department of Anatomy, Faculty of Medicine, Public Health and Nursing, Universitas Gadjah Mada, Farmako St., Sekip Utara, Senolowo, Sinduadi, Mlati, Sleman regency, Yogyakarta 55281

[3]Department of Pharmacology and Therapeutics, Faculty of Medicine, University of Indonesia Salemba Raya St. No. 6, Jakarta 10430

## ABSTRACT

The purpose of this study was to obtain natural drugs from brown seaweed (*Sargassum crassifolium*) as antiatherosclerosis candidates through the study of hypolipidemic mechanisms of action. Modeling of dyslipidemia rats was carried out by feeding high-fat (HFF) and doses of crude fucoidan 100. 200. 400mg / KgBB. in both treatments measured blood lipid profile levels taken from the orbital sinuses. HE's histopathology. mRNA expression. immunohistochemistry (IHC) with parameters VCAM-1. ICAM-1. and MCP-1 were performed on adipose tissue. as well as liver. Total cholesterol values 51.07-225.2. triglycerides 30.43-115.73. HDL 13.1-24.86 mg/dl and LDL 20.22-189.68 mg/dl. In the treatment of crude fucoidan obtained the result of p value < α (0.05. Histopathological features of adipose tissue after administration of HFF for 60 days resulted in an increase in adipose cell size. and the liver experienced structural damage and inflammation. but after 21 days of treatment the morphological picture of adipose tissue was similar to normal morphology and the liver also decreased in severity and inflammation. The results of histochemical staining after treatment showed a positive staining part on MCP-1. The result of p value < α (0.05) of mRNA expression for administration of 3 treatment doses. A dyslipidemic mouse model with HFF administration for 60 days succeeded in becoming a dyslipidemic rat. and crude fucoidan had hypolipidemic activity. Doses of 100. 200. and 400 mg/KgBB crude fucoidan showed improvement in adipose and liver morphological features of severity and inflammation of dyslipidemic rats and decreased mRNA expression.

## INTRODUCTION

Hypercholesterolemia is a risk factor for coronary heart disease (CHD) which ranks first in death in Indonesia. There are about 36 million people or about 18% of the Indonesian population who suffer from this disorder characterized by high cholesterol levels reaching values of more than 240 mg/dl. Some factors that can put a person at high risk of hypercholesterolemia are obesity. high-acid foods. low-fiber foods. lack of physical activity and smoking. Hypercholesterolemia is a major factor in atherosclerosis that causes sudden blockage of blood flow and there is a fact that weight gain is associated with an increased prevalence of dyslipidemia. diabetes mellitus. and hypertension[1].

Several epidemiological studies have identified dyslipidemia conditions which are one of the risk factors for cardiovascular problems. which play a role in the occurrence of vascular endothelial dysfunction processes[2]. Endothelial dysfunction is characterized by a decrease in the ability of the endothelium to carry out homeostasis functions such as cell migration in adhesion and activation of leukocytes. as well as immunological and inflammatory processes. This condition can cause impaired vascular tone for vasodilation and vasoconstriction. as well as conditions referred to as endothelial activation. namely pro-inflammatory. proliferative and procoagula. The pathophysiology of endothelial dysfunction is complex and involves many mechanisms. Dysfunctional endothelium is different from healthy endothelium. In endothelial dysfunction there is an increase in pro-inflammatory biomarkers both VCAM-1. ICAM-1. and MCP-1. where these three mediators can be used as molecular markers of endothelial damage[3].

Dyslipidemia therapy is mostly aimed at reducing symptoms or affecting hemodynamic responses and often does not affect the cause or course of the disease[4]. One of the anticholesterol is the statin class which is an anticholesterol drug that is widely used because of the effects of hypolipidemia. The effect of treatment with statins is estimated to be successful in achieving the target level *of low density lipoprotein* (LDL) cholesterol. but not to regression of atherosclerosis lesions[5]. The process of dyslipidemia can develop over many years. so therapy must be a long-term or even lifelong treatment. including dyslipidemia due to obesity[2]. Dyslipidemia therapy must also be followed by healthy lifestyle changes. both exercise and low-cholesterol foods. This is often difficult to do so that even though you have used antidyslipidemia drug therapy. the condition of dyslipidemia is still not corrected properly. For this reason. it is necessary to carry out complementary therapies in conjunction with conventional medicine. where this can be done by developing drugs of natural origin both on land and at sea which are used as complementary therapies[4].

Brown seaweed (*Sargassum polycystum*) is one of the natural ingredients of the sea that is used as medicine and other benefits for health. Sargassum has benefits including potential antiviral and cytotoxins[6]. as an antidiabetic by inhibiting the activity of alpha-amylase and alpha-glucosidase enzymes[7] and increasing Nfkb and IL6 expression in streptozotocin-induced diabetic rats[8]. Sargassum is also reported to protect rats from lipid peroxidation caused by acetaminophen [9] and decreased ankle joint edema of rats suffering from arthritis[10].

One of the efficacious chemical ingredients of *Sargassum polycystum* is fucoidan and in *vitro* proven to have efficacy as an anticoagulant. antidyslipidemia[11]. Fucoidan is a group of sulfate polysaccharides found in the cell wall of brown seaweed (*Sargassum polycystum*).

containing mostly L-fucosa sulfate with small amounts of monosaccharides such as galactose. glucose. xylose uronic acid[12]. Various activity studies of crude Fucoidan as a cardiovascular drug candidate have been conducted stating that crude Fucoidan has antiplatelet activity by prolonging bleeding and coagulation time and decreasing plasma uptake in rats[13]. Crude Fucoidan from *Sargassum crassifolium* extract has anti-inflammatory activity through inhibition of ICAM-1 and VCAM-1 levels in RAW 264.7 cells induced with lipopolysaccharides[14]. So this study aims to see the effect of Sargassum extract in reducing lipid profiles in vitro.

## MATERIAL AND METHODS

**Materials**

Brown seaweed (*Sargassum polycystum*), a high-fat feed made from (corn flour 2%, vitamins and minerals 2%, Na CMC 1%, sucrose 16%, casein 10%, quail egg yolk 46%, and animal fat 23%), aquadest. demineralized aquadest, CaCl2, Ethanol 96%, Ethanol 30%, and Ethanol 70%, RNA triazol solution. ethanol 70% in DEPC, pure DEPC, OligoDT, DEPC-treated H2O, RT-Buffer, DEPC-treated $H_2O$, RTase enzyme. The primers used for PCR are GAPDH. VCAM-1, ICAM-1, MCP-1, Primer-BLAST NCBI gene primers. The reagent kit used for PCR is Promega GoTaq® Green Master Mix, Gel electrophoresis 2%, Agarose, TBE gel staining, CHOD PAP cholesterol kit, triglycerides GPO method, HDL-cholesterol PTA precipitant. under the brand Biolabo Reagents, Male white rats of the wistar strain are 2-3 months old with a body weight of 150-300 grams. Tools: Spectrophotometer (Microlab 300*). Transmission Electron Cryomicroscopy (TEM) (JEM-1400 JEOL, USA), CO2 incubator, Bio Safety Cabinet Class II (Labconco*. Kansas. USA), inversion microscope, microculture plate 6 and 96 well (Iwaki, Germany), mikropipette (USA), petri disk (Iwaki®, Germany), sterile conical tube 15 ml (Iwaki® Germany). microtube 1.5 ml (Eppendorf. Germany). micropipette 10 μl and 100 μl (Nichipet Premium), microtube rack, micro centrifuge (Thermo Scientific. Germany). PCR machine (SelectCyclerTM, USA), PCR tube 0.2 ml (Axygen®. USA), electrophoresis device. gel documentary (gel doc) (Glite 600 UV Gel Documentation System, USA).

**Methods**

**Preparation of extract**

Brown Seaweed (*Sargassum polycystum*) obtained from the beach in Garut. West Java. It used as a sample has been determined in the Biota Collection Room of the University of Indonesia. Depok. West Java. Indonesia. a total of 250 g of Fine powder of brown seaweed (*sargassum sp*) macerated with 80% ethanol solvent (Pretreatment Solvent) for 2 × 12 hours at room temperature, then left overnight, after which filtered using filter paper and then taken the residue into brown seaweed ethanol extract residue. The residue of brown seaweed ethanol extract is then dried, then macerated with CaCl2 at a temperature of 25ºC for 1 × 6 hours, followed by filtration, evaporated in a rotary evaporator, and thickened until a thick extract is obtained.

**Preparation of dyslipidemic rats**

A total of 36 male white rats (*Rattus norvegicus*) of the wistar strain aged 2-3 months with body weight (150-170 grams) were induced high-fat diet (HFD) consisting of 2% corn flour, 2% vitamins and minerals, 1% Na CMC, 16% sucrose, 10% casein, 46% quail egg yolk, and 23% animal fat and given at 08.00 am and noon every day for 60 days orally with controlled conditions. All experimental animals were then measured lipid profiles which included total cholesterol, triglycerides, LDL, and HDL levels until dyslipidemia was confirmed. All test protocols have been approved by the ethics commission with letter numbers.

**The antidyslipidemic test to animal model**

The animals model were then divided into 6 groups ant treated with:

| | |
|---|---|
| Group Normal | : Normal rats with distilled water 0.2 mL/day |
| Group Negative | : Dyslipidemic rats with distilled water 0.2 mL/day |
| Group Dosage 1 | : Dyslipidemic rats treated with Fucoidan Dosage 100mg/KgBW/day |
| Group Dosage 2 | : Dyslipidemic rats treated with Fucoidan Dosage 200mg/KgBW/day |
| Group Dosage 3 | : Dyslipidemic rats treated with Fucoidan Dosage 400mg/KgBW/day |
| Group Positive Control | : Dyslipidemic rats treated with Simvastatin dose 10 mg/KgBW/day |

The treatment of crude fucoidan was given to dyslipidemic rats for 21 days orally and then at the end of the study assessed changes in lipid profile parameters: total cholesterol levels, triglycerides, LDL, and HDL. Blood was collected from the tail vein and the parameters was estimated using GOD-PAP reagen kit with spectrophotometric method. The animals were sacrificed by cervical decapitation after anasthesia. The adipose tissue and liver dyslipidemic rats were collected to identifying histopathological features and mRNA expression by assessing VCAM-1. ICAM-1 and MCP-1 expression.

## RESULT AND DISCUSSION

**Plant determination**

The results of the determination showed that the plant used was (*Sargassum polycystum*). The extract obtained has a dark brown color. a characteristic smell of fishy seaweed with a yield of 11.68%. DER-Native 8.56%. Acidity (pH) 6.7. soluble in water and ethanol. very soluble in chloroform. The yield of crude Fucoidan extracted with a 2% $CaCl_2$ solution was obtained at 11.07%. where the extraction with this method was in accordance with the soaking value obtained around 5-10% from previous studies. The organoleptic results of crude Fucoidan are brown because brown seaweed contains pigments. namely fucosanthine. but the distinctive smell of brown seaweed was originally fishy. lost this because it has been treated with the use of 80% ethanol which is useful for separating salt. fat. residues from seawater. The preparation of *Sargassum polycystum powder is* processed by freezedrying method at a temperature of -40 ºC. pressure 37-133 Pa for 60 hours and Fucoidan crude pH obtained 6.68 which is a neutron pH. this is due to the use of $CaCl_2$ solution and pure water that has a neutral pH in the extraction process.

**Measurement of lipid level parameters after high-fat feeding**

Lipid levels were measured on days 0 and 60. which obtained the following results:

**Table 1. Lipid profile before high-fat feed (Day 0)**

| Group | Lipid Profile ± SD (mg/dL) | | | |
|---|---|---|---|---|
| | **Cholesterol** | **Triglycerides** | **HDL** | **LDL** |
| **Control** | 44.95±5.26 | 19.15±5.003 | 31.18±1.91 | 9.94±5.29 |
| **HFF** | 52.27±4.06 | 25.25±11.72 | 33.17±2.33 | 14.05±6.5 |
| **Simvastatin** | 47.07±4.94 | 22.00±4.86 | 32.38±2.13 | 10.28±3.80 |
| **Dose 1** | 47.20±5.72 | 19.75±5.36 | 30.88±2.44 | 12.37±4.98 |
| **Dose 2** | 44.02±8.91 | 23.40±7.09 | 27.93±7.57 | 11.41±4.20 |
| **Dose 3** | 45.76±2.35 | 19.53±2.25 | 30.70±1.2 | 11.16±3.97 |

The data mentioned above showed that before feeding high-fat lipid levels entered the normal range of rats. namely total cholesterol 10-54 mg/dl. triglycerides 27.89-29.44 mg/dl. HDL 34.1-35.5 mg/dl and LDL 7-27.2 mg/dl. It shows the average range of results. namely total cholesterol 44.02-52.27 mg/dL. triglycerides 19.15-25.25 mg/dL. HDL 27.93-33.17 mg/dL and LDL 9.94-14.05 mg/dL.

**Table 2. Lipid profile after high-fat feed (Day 60)**

| Group | Lipid Profile ± SD (mg/dL) | | | |
|---|---|---|---|---|
| | **Cholesterol** | **Triglycerides** | **HDL** | **LDL** |
| **Control** | 51.07 ± 3.3 | 30.43 ± 3.2 | 24.86± 1.5 | 13.5 ± 1.6* |
| **HFF** | 196.56± 11.2* | 109.36 ± 12.9* | 15.48 ± 4.0 | 159.21 ± 11.2* |
| **Simvastatin** | 205.53 ± 7.0* | 112.1 ± 9.1* | 14.13 ± 2.6 | 121.60 ± 6.4* |
| **Dose 1** | 208.75± 18.2* | 111.13 ± 11.1* | 17.65 ± 3.6 | 155.38 ± 24.7* |
| **Dose 2** | 225.2 ± 29.4* | 112.13 ± 10.0* | 13.1 ± 2.8* | 189.68± 28.2* |
| **Dose 3** | 193.3 ± 7.0* | 105.5 ± 13.2* | 15.16 ± 3.6 | 157.03 ± 6.8* |

The data in the table above showed that after feeding high-fat feed for 60 days there was an increase in lipid profile levels for total cholesterol. The triglycerides and LDL exceeding normal range levels of healthy mice. but for HDL levels occurred lower than normal range levels of mice. On day 60 when compared to day 45. there was a significant increase and decrease in lipid profile levels compared to the control group (normal rat group). Based on the results of lipid profile levels on day 60. it can be determined that on day 60 there has been a rats model of dyslipidemia. Average total cholesterol 51.07-225.2 mg/dL. triglycerides 30.43-112.13 mg/dL. HDL 13.1-24.86 mg/dL and LDL 13.5-189.68 mg/dL.

**Antidyslipidemia activity test results of brown seaweed (*Sargassum polycystum*)**

The observation of total cholesterol. triglycerides. HDL and LDL was carried out after administration of fucoidan extract after 21 days as if given treatment with control.

*Total cholesterol level*

The total cholesterol levels were measured after crude fucoidan treatment for 21 days in dyslipidemia model rats (60 days HFF administration)

**Table 3. Total cholesterol levels after 60 days of HFF and fucoidan for 21 days**

| Group | Average of total cholesterol levels (mg/dL) | | |
|---|---|---|---|
| | H0 | H60 | H81 |
| Control | 44.95 ± 5.6 | 51.07±3.4 | 50.5±1.8** |
| HFF | 52.27 ± 4.1 | 205.53±7.1* | 211.8±5.9 |
| Simvastatin | 47.07 ± 4.9 | 196.56±11.2* | 55.4±2.2** |
| D1 | 47.2 ± 5.7 | 208.75±18.2* | 64.4±53.7** |
| D2 | 44.02 ± 8.9 | 225.2±29.4* | 55.2±43.3** |
| D3 | 45.76 ± 2.4 | 193.3±7.0* | 54.7±1.08** |

This result shows the activity of giving crude fucoidan dosage of 1,2 and 3 (100, 200, 400 mg/Kg BW/day) for 21 days reduce total cholesterol levels in dislipdemia rats given high-fat feed and significantly different to control (p<0.001) and percentage reduction can be seen in Fig 1.

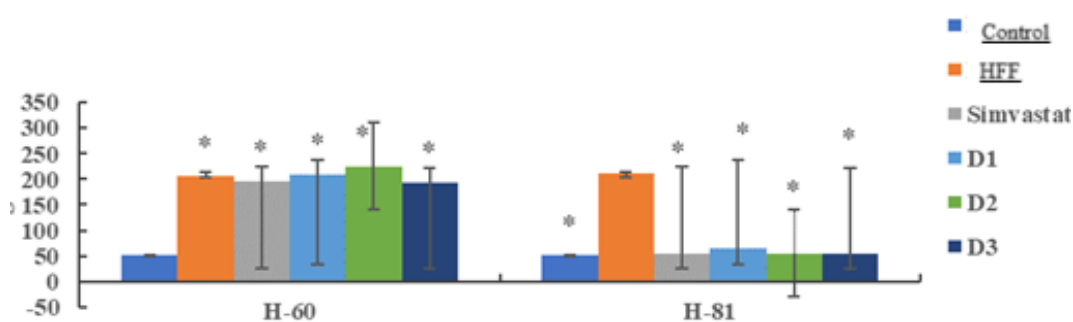

**Fig 1. Total cholesterol levels after high fat feed after 60 days and fucoidan 21 days**

Based on the results of the data above can be seen that there was a decrease in total cholesterol levels in the positive group by 76.49% of the dose group also experienced a decrease in cholesterol levels by 51.05% at a dose of 100 mg/KgBW/day 69.14% at a dose of 200 mg/kgBW/day and 75.49% at a dose of 400mg/kgBB/day. Analysis of the data obtained from the normality test. namely normal distributed data. then continued with the homogeneity test obtained results. namely homogeneous distributed data. then a one-way ANOVA test was carried out obtained p < 0.01 which means there is a significant difference then a post hoc BNT / LSD test was carried out which means there is a significant difference between the negative group and the normal group. positive and dosage. The positive group also had significant differences in the dose groups of 100, 200, and 400 mg/KgBB/day.

*HDL levels*
HDL levels were measured after crude fucoidan treatment for 21 days in dyslipidemia model

**Table 4. HDL levels after 60 days of HFF and Crude Fucoidan administration for 21 days**

| Group | Days to- | | | | | | % day 0-60 | Result | % day 60-81 | Result |
|---|---|---|---|---|---|---|---|---|---|---|
| | 0 | 14 | 28 | 45 | 60 | 81 | | | | |
| Control | 31 | 30.9 | 29.18 | 29.03 | 24.86 | 24.2 | 20.27 | Decreased | 2.65 | Decreased |
| Simvastatin | 32 | 27.1 | 27 | 23.38 | 15.48 | 31.1 | 52.19 | Decreased | 50.23 | Increased |
| Negative | 33 | 19.8 | 17.88 | 14.3 | 14.13 | 13.9 | 57.40 | Decreased | 1.63 | Decreased |
| Dosage 1 | 31 | 26.2 | 23.92 | 19.43 | 17.65 | 18.6 | 42.84 | Decreased | 5.11 | Increased |
| Dosage 2 | 28 | 22.4 | 20.13 | 16.76 | 13.1 | 21.6 | 42.84 | Decreased | 39.35 | Increased |
| Dosage 3 | 31 | 28.7 | 23.86 | 20.06 | 15.16 | 29.5 | 53.10 | Decreased | 48.61 | Increased |

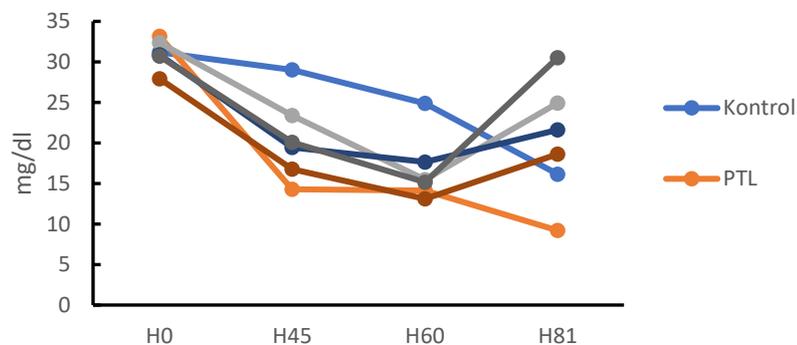

**Fig 2. HDL levels after HFF for 60 days and Crude Fucoidan for 21 days**

Based on the results above. data were obtained that there was an increase in HDL levels in the positive group by 37.83% (Table 4). The dose group also experienced an increase in HDL levels by 47.47% at the 100mg/KgBB dose, 18.29% at the 200mg/KgBB dose, and 29.57% at the 400mg/KgBW dose. Analysis of data obtained from the normality test. namely normal distributed data. then continued with the homogeneity test obtained results. namely homogeneous distributed data. Furthermore. a one-way ANOIVA test was carried out with $p < α$ results (0.05) which means there is a significant difference then a post hoc BNT / LSD test was carried out which means there was a significant difference in the dose group with the negative group and the positive group found a significant difference in the dose group of 100, 200, and 400 mg/ KgBW.

*LDL levels*

LDL levels were measured after crude fucoidan treatment for 21 days in dyslipidemia model mice.

Tabel 5. 11. LDL levels after Crude Fucoidan treatment for 21 days

| Group | Days to | | | | | | % 0-60 | Result | % 60-81 | Result |
|---|---|---|---|---|---|---|---|---|---|---|
| | 0 | 14 | 28 | 45 | 60 | 81 | | | | |
| Control | 9.94 | 12.65 | 14.77 | 14.93 | 13.5 | 14 | 26.37 | Increased | 3.60 | Increased |
| Simvastatin | 10.28 | 60.13 | 79 | 96.81 | 159.21 | 15.62 | 93.54 | Increased | 90.20 | Decreased |
| HFD | 14.05 | 71.5 | 87.91 | 112.1 | 121.6 | 127.2 | 88.45 | Increased | 3.40 | Increased |
| Dosage 1 | 20.86 | 79.2 | 89.26 | 127.5 | 171.65 | 35.65 | 87.85 | Increased | 78.65 | Decreased |
| Dosage 2 | 12.37 | 65.04 | 83.38 | 121.3 | 155.38 | 26 | 92.04 | Increased | 83.27 | Decreased |
| Dosage 3 | 11.41 | 75.13 | 87.93 | 127.1 | 189.68 | 18.4 | 93.98 | Increased | 90.3 | Decreased |

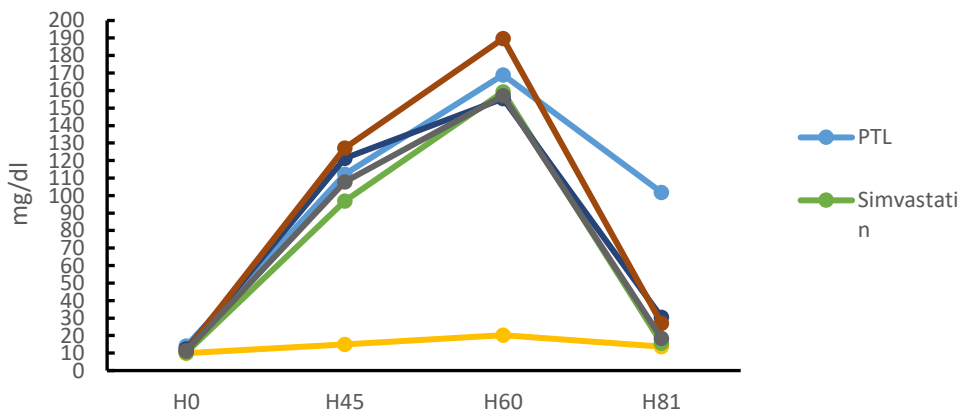

**Fig 3. LDL levels after administration Crude Fucoidan for 21 days.**

Based on the results above. data were obtained on a decrease in LDL levels in the positive group by 90.20% (table IV.12). The dose group also experienced a decrease in LDL levels by 65.34% at 100mg/KgBB. 44.39% at 200mg/KgBB and 59.07% at 400mg/KgBB. Analysis of data obtained from the normality test. namely normal distributed data. then continued with the homogeneity test obtained results. namely homogeneous distributed data. Furthermore. a one-way ANOIVA test obtained $p < \alpha$ results (0.05) which means there is a significant difference then a post hoc BNT / LSD test was carried out which means there was a significant difference in the dose group with the negative group and the positive group obtained a significant difference in the dose group of 100. 200. and 400 mg / KgBB.

White rats are used because they have beneficial properties. namely rapid breeding. have a larger size than mice. are easy to maintain in large quantities and are easily adapted in a laboratory environment. and most importantly in this study required a considerable blood volume sample of 1-2 mL. which was used for measurement of lipid profile parameters in dyslipidemic animal model mice. Increased total cholesterol levels can be caused by the intake of high-fat diet foods Egg yolks and goat oil become sources of animal cholesterol and fat that can increase cholesterol levels because they are high in cholesterol and saturated fatty acids. The yolk has the highest fat content, Total cholesterol, triglycerides and LDL levels are quite high because every consumption of saturated fat as much as 1% of total energy a day is thought to increase cholesterol levels to 2.7 mg/dL (68).

Based on the results mentioned above. it was seen that there was an increase in total cholesterol. triglycerides. LDL and a decrease in HDL seen in the measurement of lipid profile levels on days 0, 14, 28, 45, and 60 after high-fat feeding. namely the presence of cholesterol obtained from quail eggs because quail eggs have higher cholesterol levels (844 mg/dl) compared to chicken egg cholesterol levels (423 mg/dl) so that they can improve lipid profile levels better compared to using chicken eggs. Animal fat is a saturated fatty acid so it triggers an increase in total and LDL cholesterol levels if consumed continuously (69). Giving therapy such as simvastatin can reduce total cholesterol so that there are differences between positive groups and other groups. The dose group was also given crude fucoidan therapy of brown seaweed with different doses. namely doses of 100. 200 and 400 mg / KgBB. Giving crude fucoidan brown seaweed can reduce total cholesterol due to the content of Fucoidan. fucosanthin and alginate. Cholesterol levels can decrease due to LCAT (*Lecithin Cholesterol Acyltransferase*) activity and can lower lipid levels by activating serum LPL and HL activity (70).

Giving brown seaweed can lower cholesterol and can also increase HDL levels although not significantly. Increasing dosing can affect HDL levels can be seen from the graph of an increase in HDL levels at the third dose. which is 200 mg / KgBB. The compound content in brown seaweed. namely fucosanthin. can activate SR-B1 so that fucosanthine can suppress the absorption of LDL and HDL from the blood to the liver. Giving therapy with brown seaweed can also reduce LDL levels. The level of LDLR expression in the liver is greatly decreased due to fucosanthine. Reduced LDLR and SR-B1 expression levels in the liver suggest that fucosanthine suppresses the absorption of LDL particles (70).

*Histopathological Features of Hepatic Tissue Hematoxilin and Eosin Staining (H&E)*

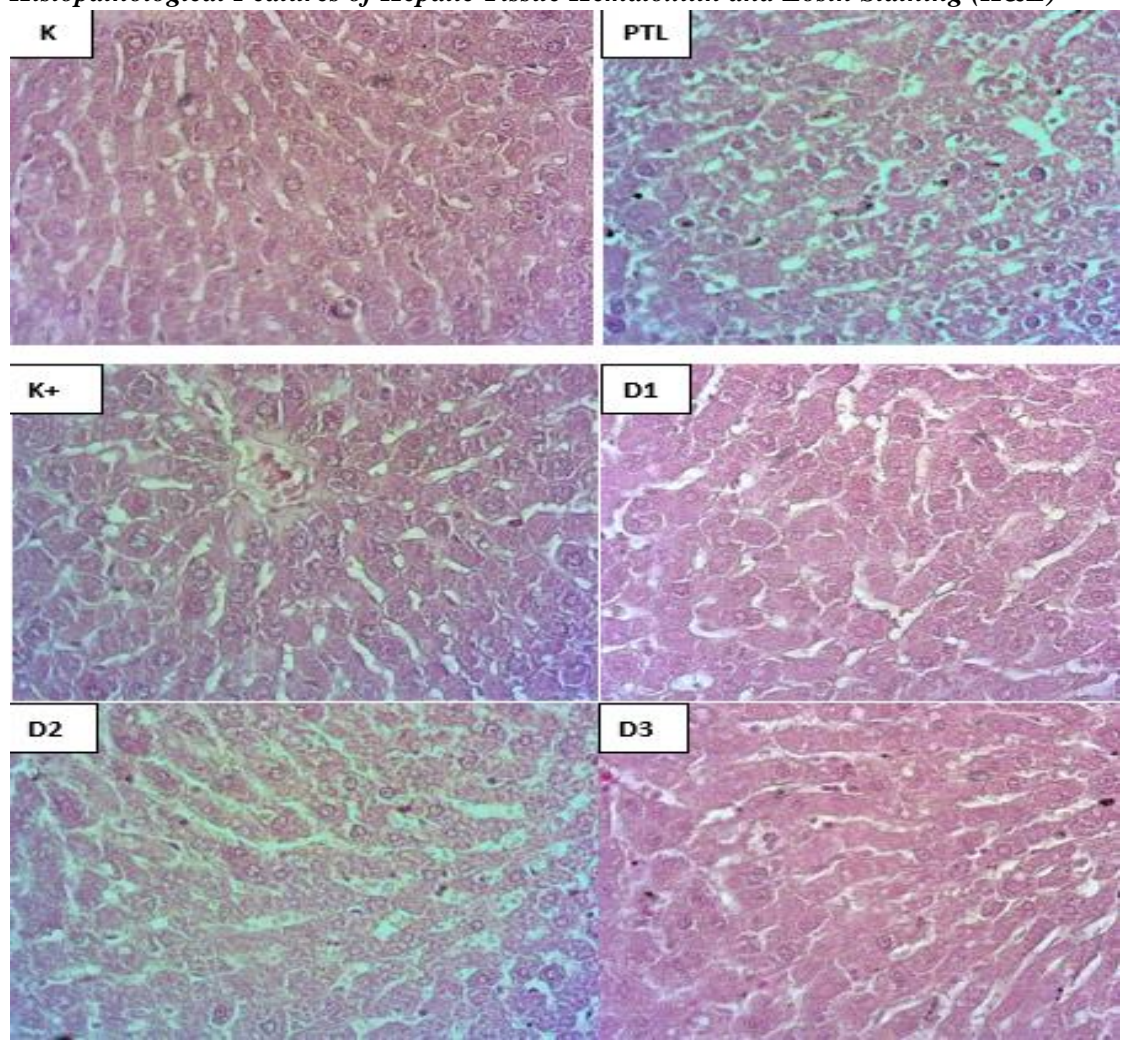

**Figure 4. Results of transverse hepatic histopathological staining using H&E staining (400x). K=Control Group. HFF= Negative Control Group (HFF). K+=Positive Control Group (Simvastatin). D1=Dose 1 (100 mg/kgBB). D2= Dose 2 (200 mg/kgBB). D3= Dose 3 (400 mg/kgBB).**

The figure 4 above is the result of histopathological structure in the liver of dyslipidemic rats which was also seen using H&E staining. it was seen that the liver cell structure in the normal control group showed normal and clean conditions without fat deposition and inflammation. The picture of histopathological structures in the liver of the negative group. namely those given high-fat feed alone causes fat deposition in hepatocyte cells. the structure of the liver is damaged and inflammatory. However. in the crude fucoidan dosing group. the morphological structure of liver cells decreased slightly in severity and inflammation.

*VCAM-1 mRNA expression in the liver*

VCAM-1 mRNA expression was performed in mouse livers with a dyslipidemia model. The bands formed from electrophoresis results were then quantified by densitometric analysis on ImageJ. In the liver. the obtained expression values were tested for normality using the Saphiro Wilk test because the samples numbered less than 50. The test results showed that mRNA expression in each treatment group was normally distributed (P>0.05). The *Homogeneity of Variances* value obtained VCAM-1 significance value for the liver is 0.164 which indicates that the value of mRNA expression variance is statistically different (P=0.05). because the data obtained are normally distributed. the *One Way ANOVA test can be performed*.

The test results. for the liver obtained a value of P<0.05 (P = 0.000) which showed there was a significant difference in the treatment group. To find out which groups have significant differences in mRNA expression values. then proceed with *Post Hoc Tests*. the results of analysis for VCAM-1 mRNA in the liver showed that administering 4 doses of treatment to mice was able to significantly reduce VCAM-1 expression compared to HFF control (p = 0.000).

The results of VCAM-1 mRNA expression analysis in the liver with several treatments statistically can be seen in figure- below and appendix18.

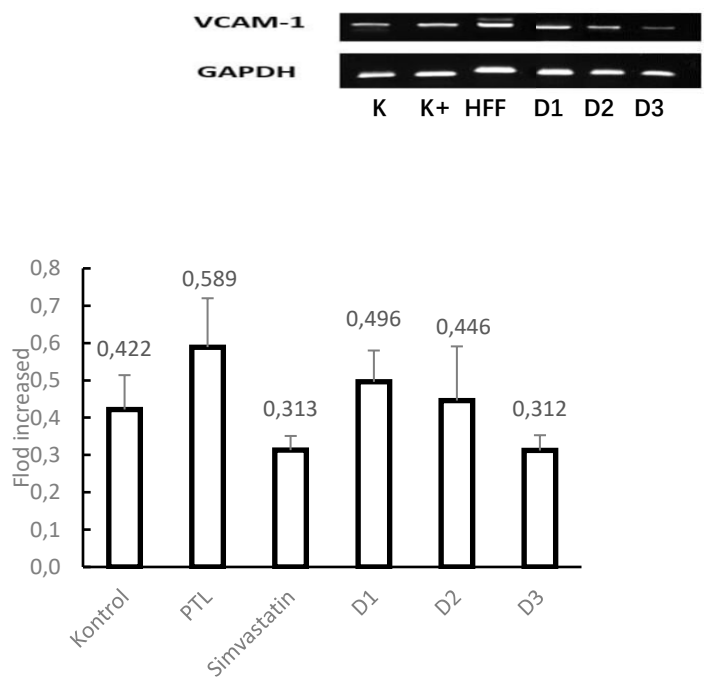

**Figure 5. Electrophoresis Band and VCAM-1 mRNA Expression Graph in mouse liver dyslipidemia model analyzed by RT-PCR with 6 different treatments. Data is presented in mean±SD. *P<0.05 versus the HFF group.**

*ICAM-1 mRNA expression in the liver*

ICAM-1 mRNA expression was performed in mouse livers with dyslipidemia models. The bands formed from electrophoresis results were then quantified by densitometric analysis on ImageJ. In the liver. the obtained expression values were tested for normality using the Saphiro Wilk test because the sample numbered less than 50. The test results showed that mRNA expression in each treatment group was normally distributed (P>0.05). for the *Homogeneity of Variances value.* the significance value of ICAM-1 for the liver is 0.293 which indicates that the value of mRNA expression variance is statistically different (P>0.05). The data obtained is normally distributed. so the *One Way ANOVA* test can be carried out. The test results. for the liver obtained a value of P<0.05 (P = 0.000) which showed there was a significant difference in the treatment group. To find out which groups have meaningful differences in mRNA expression values. then proceed with *Post Hoc Tests*. The results of the analysis for ICAM-1 mRNA in the liver showed that dosing 2. 3 and dose 4 in mice was able to significantly reduce ICAM-1 expression compared to HFF controls (p = 0.003; 0.000; 0.000). Here are the results below and appendix 18:

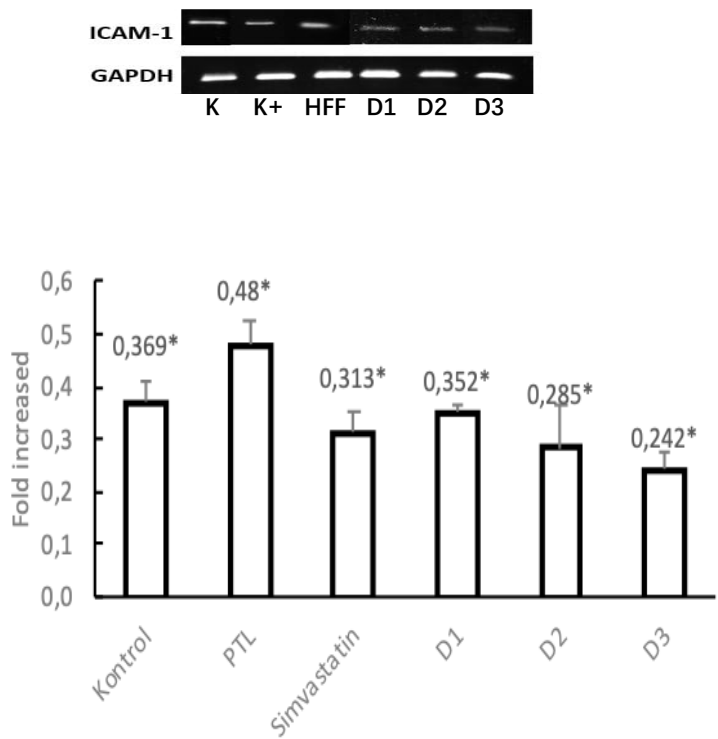

**Figure 6. Electrophoresis Band and ICAM-1 mRNA expression graph in rat liver dyslipidemia model analyzed by RT-PCR with 6 different treatments. Data is presented in mean±SD. *P<0.05 versus HFF control.**

*MCP-1 mRNA expression in Liver Tissue*

MCP-1 mRNA expression was performed in the liver with a dyslipidemia model. The bands formed from electrophoresis results were then quantified by densitometric analysis on ImageJ. The obtained expression values were tested for normality using the Saphiro Wilk test because the sample numbered less than 50. The test results showed that mRNA expression in each treatment group was normally distributed ($P>0.05$). then for the *Homogeneity of Variances value obtained* the significance value of MCP-1 for the liver was 0.187 which indicated that the value of mRNA expression variance was statistically different ($P>0.05$). The results obtained are normally distributed. so the *One Way ANOVA* test can be carried out. From the test results. for the liver obtained a value of $P < 0.05$ ($P = 0.000$) which showed there was a significant difference in the treatment group. To find out which groups have meaningful differences in mRNA expression values. then proceed with *Post Hoc Tests*. Analysis results for MCP-1 mRNA in the liver showed that dosing doses 2. 3 and 4 significantly decreased MCP-1 expression compared to HFF controls ($p = 0.000; 0.000; 0.000$).

The results of liver MCP-1 mRNA expression analysis in several treatments can be statistically seen -the following figure and appendix 18:

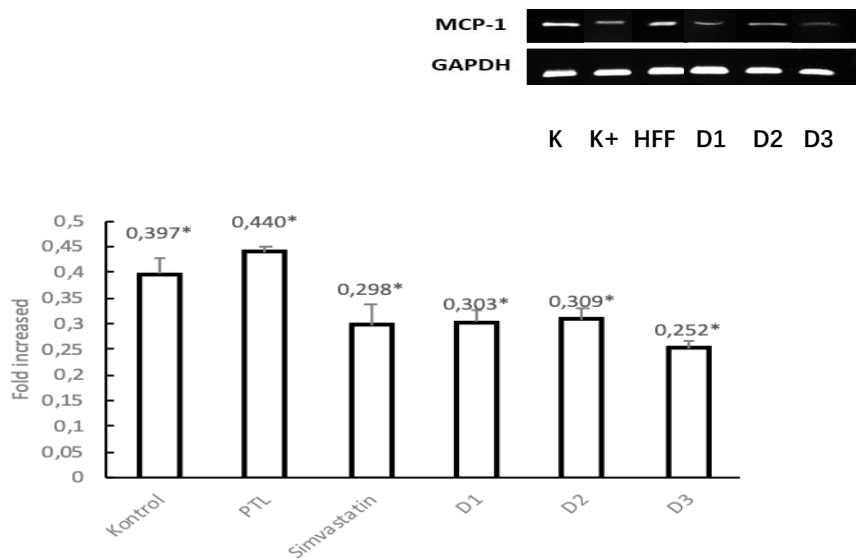

**Figure 7. Electrophoresis band and MCP-1 mRNA expression in rat liver dyslipidemia model analyzed by RT-PCR with 6 different treatments. Data is presented in mean±SD. *$P<0.05$ versus HFF control.**

*Immunohistochemistry of MCP-1 in liver Tissue*

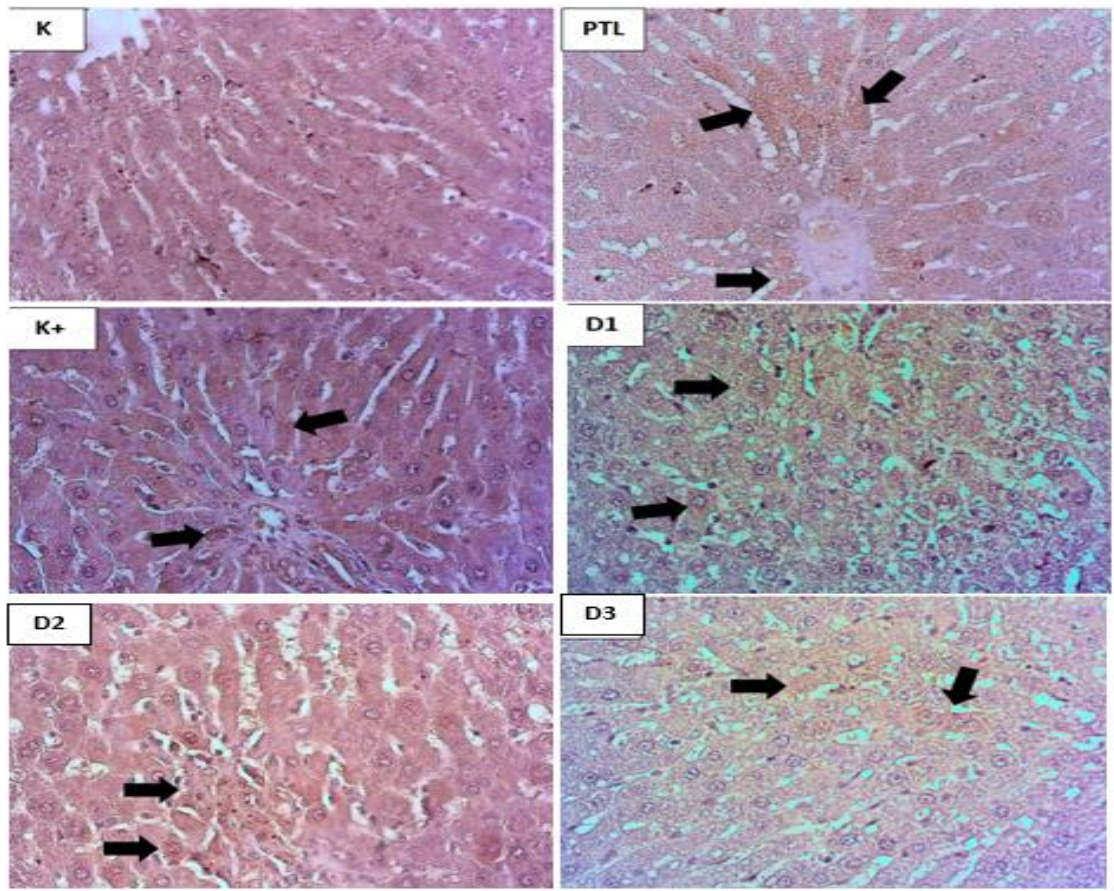

**Figure 8. Microscopic image of IHC MCP-1 in the liver transversely (400x). Black arrows indicate the positive coloring part of MCP-1. . K=Control Group. HFF= Negative Control Group (HFF). K+=Positive Control Group (Simvastatin). D1=Dose 1 (100 mg/kgBw). D2= Dose 2 (200 mg/kgBw). D3= Dose 3 (400 mg/kgBw).**

Figure 8 show of MCP-1 expression in the liver where in IHC staining the negative control rat group given only HFF or high-fat feed. showed positive results (brown). Brown staining in hepatocyte cells was seen to decrease with crude fucoidan at dose 1 (100 mg / kg body weight). dose 2 (200 mg / kg body weight). dose 3 (400 mg / kg body weight) compared to the negative control group.

Based on the results of the above histopathology in this study. showing the histopathological picture of adipose tissue and liver stained with H & E showed morphological changes in the group induced by high-fat feeding (HFF) with a normal control group. Morphological changes in adipose tissue and liver are seen in the presence of excess fat deposition in adipocytes and hepatocytes. Crude fucoidan administration in rats induced by high-fat feeding (HFF). showed adipose tissue and liver conditions that were almost close to adipose tissue morphology and liver normal control and positive control. Fucoidan administration can reduce fat ratio. The decrease in fat accumulation and hypertrophy of adipose tissue is confirmed by H&E staining of adipose tissue.

Excessive energy intake results in lipid deposition not only in adipose tissue but also in the liver. which can accelerate progression towards steatohepatitis. a serious complication of obesity. Wang *et al.* (2021) found that administration of Fucoidan caused a decrease in hepatic weight accompanied by a decrease in hepatrical triglyceride and cholesterol levels. Fucoidan administration also weakened liver injury in HFF-fed rats. Hepatic histological examination confirmed that hepatic lipid droplets caused by HFF administration dramatically decreased with Fucoidan administration.

## CONCLUSION

The formation of a dyslipidemia mouse experimental animal model with high-fat feed for 60 days succeeded in becoming a dyslipidemia and crude fucoidan had antidyslipidemia activity in the dyslipidemia mouse model with lipid profile parameters including total cholesterol, triglycerides. HDL and LDL. Histopathological features in rat adipose tissue after 60 days of high-fat feeding resulted in an increase in adipose cell size. and in the liver resulted in structural damage and inflammation. Doses of 100, 200, and 400 mg/KgBB of crude fucoidan for 21 days in dyslipidemic rats showed adipose morphological features similar to morphology under normal circumstances and the morphological structure of the liver also decreased in severity and inflammation.The expression of VCAM-1. ICAM-1. MCP-1 mRNA in adipose tissue and liver showed that administration of 3 treatment doses (100, 200, and 400mg/KgBB) in dyslipidemic rats was able to significantly reduce mRNA expression compared to HFF controls. Administration of crude fucoidan for 21 days can affect histopathological features in adipose tissue and liver of dyslipidemia model mice with histochemical staining (IHC) which shows positive staining part on MCP-1.

## ACKNOWLEDMENT

We would like to thank all supervisors, mentors, and professors in this research especially in the International Cooperation program in the field of Education, Research and Community Service with the Center for Botanicals and Chronic Diseases – Rutgers University-School of Environmental and Biological Sciences.

## FUNDING

This research reported in this publication was supported and funded by Fogarty International Center of the National Institutes of Health under Award Number D43TW009672.

# CONFLICT OF INTEREST

The authors declare no conflict of interest.